\documentclass[sigconf]{acmart}

\AtBeginDocument{%
  \providecommand\BibTeX{{%
    \normalfont B\kern-0.5em{\scshape i\kern-0.25em b}\kern-0.8em\TeX}}}

\setcopyright{acmcopyright}
\copyrightyear{2022}
\acmYear{2022}
\setcopyright{acmcopyright}\acmConference[WSDM '22]{Proceedings of the Fifteenth ACM International Conference on Web Search and Data Mining}{February 21--25, 2022}{Tempe, AZ, USA}
\acmBooktitle{Proceedings of the Fifteenth ACM International Conference on Web Search and Data Mining (WSDM '22), February 21--25, 2022, Tempe, AZ, USA}
\acmPrice{15.00}
\acmDOI{10.1145/3488560.3498390}
\acmISBN{978-1-4503-9132-0/22/02}

\usepackage{amsmath}

\usepackage{amssymb,amsfonts}
\usepackage{algpseudocode}
\usepackage{graphicx}
\usepackage{textcomp}
\usepackage{xcolor}
\usepackage{multirow}
\usepackage{fontawesome}
\usepackage{paralist}

\usepackage[linesnumbered,ruled,vlined]{algorithm2e}
\usepackage{amsthm}
\theoremstyle{definition}

\newtheorem{problem}{Problem}
\newtheorem{lemma}{Lemma}
\usepackage{bm}
\usepackage{booktabs}
\usepackage{stmaryrd}
\usepackage{comment}
\usepackage{hyperref}
\usepackage{epsfig} 
\usepackage{balance}

\usepackage[caption=false,font=normalsize,labelfon
t=sf,textfont=sf]{subfig}

\usepackage[export]{adjustbox}
\usepackage{enumitem} 
\usepackage{makecell}
\usepackage{tikz}
\usepackage{url}
\usepackage{tablefootnote}

\newtheorem{thm}{Theorem}

\newcommand{\method}{\textsc{CutNPeel}\xspace}

\newcommand{\naive}{\textsc{Peel}\xspace}
\newcommand{\naivenosep}{\textsc{Peel}}

\newcommand{\timecrunch}{\textsc{TimeCrunch}\xspace}
\newcommand{\comtwo}{\textsc{Com2}\xspace}
\newcommand{\vog}{\textsc{VoG}\xspace}
\newcommand{\mzoom}{\textsc{M-Zoom}\xspace}
\newcommand{\dcube}{\textsc{D-Cube}\xspace}
\newcommand{\slashburn}{\textsc{SlashBurn}\xspace}

\newcommand{\dynamicGSum}{\Glong}

\newcommand{\srcset}{\mathcal{\tilde{S}}}
\newcommand{\dstset}{\mathcal{\tilde{D}}}
\newcommand{\timeset}{\mathcal{\tilde{T}}}
\newcommand{\edgeset}{\mathcal{\tilde{E}}}

\newcommand{\GG}{\mathcal{G}}

\newcommand{\Glong}{\mathcal{G}=(\mathcal{G}^{(t)})_{t\in \mathcal{T}}}

\newcommand{\GT}{\mathcal{G}^{(t)}=(\mathcal{S}^{(t)}\cup\mathcal{D}^{(t)}, \mathcal{E}^{(t)})}
\newcommand{\GTT}{\mathcal{\tilde{G}}=(\srcset\cup \dstset, \edgeset)}

\newcommand{\GTlong}{\mathcal{\tilde{G}}=(\mathcal{\tilde{G}}^{(t)})_{t\in \timeset}}

\newcommand{\GTTTshort}{\mathcal{\tilde{G}}^{(t)}}

\newcommand{\E}{\mathcal{E}}
\newcommand{\ST}{\mathcal{S}^{(t)}}
\newcommand{\DT}{\mathcal{D}^{(t)}}
\newcommand{\ET}{\mathcal{E}^{(t)}}

\newcommand{\vi}{s}
\newcommand{\vj}{d}

\newcommand{\edge}{e=(s,d,t)}
\newcommand{\SSS}{\mathcal{S}}
\newcommand{\D}{\mathcal{D}}

\newcommand{\nearset}{\mathcal{\hat{B}}}
\newcommand{\blockset}{\mathcal{B}}

\newcommand{\IT}{\mathcal{\tilde{I}}}

\newcommand{\T}{\mathcal{T}}
\newcommand{\I}{\mathcal{I}}

\newcommand{\GIT}{\mathcal{G}_{\IT}}
\newcommand{\EIT}{\mathcal{E}_{\IT}}
\newcommand{\EPIT}{\mathcal{E}_{\IT}'}
\newcommand{\EITM}{\mathcal{E}_{\IT \setminus \{i\}}}

\newcommand{\PIT}{\blockset_{\mathcal{\tilde{I}}}}
\newcommand{\PITP}{\blockset_{\mathcal{\tilde{I}'}}}
\newcommand{\PITM}{\blockset_{\IT \setminus \{i\}}}

\newcommand{\GPIT}{\mathcal{G}_{\IT}'}
\newcommand{\GP}{\mathcal{G}'}
\newcommand{\EP}{\mathcal{E}'}

\newcommand{\objF}{\phi}
\newcommand{\objB}{\phi(\nearset,\GG)}

\newcommand{\LPIT}{\mathcal{L}_{\I}(\PIT)}
\newcommand{\LPITP}{\mathcal{L}_{\mathcal{\tilde{I}}'}(\PITP)}

\newcommand{\LE}{\mathcal{L}_{e}}

\newcommand{\correctionP}{\mathcal{R}}
\newcommand{\correctionM}{\mathcal{M}}

\newcommand{\Gi}[1]{\GG^{\prime}_{\left(#1\right)}}

\newcommand{\Ei}{\E^{\prime}_{\left(k\right)}}
\newcommand{\V}{\mathcal{V}}
\newcommand{\Vi}{\I-\V\cup\V_{\left(k\right)}}
\newcommand{\Vk}{\V_{\left(k\right)}}
\newcommand{\Vgroups}{\cup_{i} \{\V_{\left(i\right)}\}}
\newcommand{\vvv}{v}

\newcommand{\Ggroups}{\cup_{i} \{\GG^{\prime}_{\left(i\right)}\}}

\newcommand{\iter}{T}

\newcommand{\decrement}{\alpha}
\newcommand{\threshold}[1]{\theta(#1)}
\newcommand{\shingleGraph}[1]{f(#1)}

\newcommand{\densityMax}{s_{max}}
\newcommand{\saving}[1]{\textit{Saving}(#1)}
\newcommand{\density}[1]{\rho(#1)}
\newcommand{\peeler}{\textsc{PeelOne}\xspace}
\newcommand{\dividing}{\textsc{Cut}\xspace}

\newcommand{\smallsection}[1]{{\vspace{0.05in} \noindent {\bf{\underline{\smash{#1}:}}}}}

\begin{document}

\title{Finding a Concise, Precise, and Exhaustive Set of Near Bi-Cliques in Dynamic Graphs}


\settopmatter{authorsperrow=4}
\author{Hyeonjeong Shin}
\affiliation{%
  \institution{KAIST}
  \city{Seoul}
  \country{South Korea}
}
\email{hyeonjeong1@kaist.ac.kr}

\author{Taehyung Kwon}
\affiliation{%
  \institution{KAIST}
  \city{Seoul}
  \country{South Korea}
}
\email{taehyung.kwon@kaist.ac.kr}

\author{Neil Shah}
\affiliation{%
  \institution{Snap Inc.}
  \city{Seattle}
  \state{Washington}
  \country{USA}
}
\email{nshah@snap.com}

\author{Kijung Shin}
\affiliation{%
  \institution{KAIST}
  \city{Seoul}
  \country{South Korea}
}
\email{kijungs@kaist.ac.kr}


\fancyhead{}
\settopmatter{printacmref=true}

\begin{abstract}
    A variety of tasks on dynamic graphs, including anomaly detection, community detection, compression, and graph understanding, have been formulated as problems of identifying constituent (near) bi-cliques (i.e., complete bipartite graphs).
Even when we restrict our attention to maximal ones, there can be exponentially many near bi-cliques, and thus finding all of them is practically impossible for large graphs. 
Then, two questions naturally arise: 
(Q1) What is a ``good'' set of near bi-cliques?
That is, given a set of near bi-cliques in the input dynamic graph, how should we evaluate its quality?
(Q2) Given a large dynamic graph, how can we rapidly identify a high-quality set of near bi-cliques in it? 
Regarding Q1, we measure how concisely, precisely, and exhaustively a given set of near bi-cliques describes the input dynamic graph. 
We combine these three perspectives systematically on the Minimum Description Length principle.
Regarding Q2, we propose \method, a fast search algorithm for a high-quality set of near bi-cliques.
By adaptively re-partitioning the input graph, \method reduces the search space and at the same time improves the search quality.
Our experiments using six real-world dynamic graphs demonstrate that \method is 
(a) \textit{High-quality}: providing near bi-cliques of up to $\mathbf{51.2\%}$ \textbf{better quality} than its state-of-the-art competitors, 
(b) \textit{Fast}: up to $\mathbf{68.8\times}$ \textbf{faster} than the next-best competitor, and (c) \textit{Scalable}: scaling to graphs with \textbf{$\mathbf{134}$ million edges}. We also show successful applications of \method to graph compression and pattern discovery.

\end{abstract}

%
%



%


\maketitle

\section{Introduction}
    \label{sec:intro}
Daily activities on the Web generate an enormous amount of data.
Especially, many activities, such as e-mail communications, web surfing, online purchases, result in data in the form of graphs, such as e-mail networks, IP-IP communication networks, and user-business bipartite graphs.
Most real-world graphs, including the aforementioned ones, evolve over time, and thus each of them is represented as a \textit{dynamic graph}, i.e., a sequence of graphs over time.

A \textit{bi-clique} is a complete bipartite graph.
That is, a bi-clique consists of two disjoint sets of nodes where every node in a set is adjacent to every node in the other set.
We use the term \textit{near bi-clique} to denote a bipartite graph ``close" to a bi-clique with few or no missing edges.
We intentionally leave the concept near bi-clique flexible without rigid conditions (e.g., conditions in \cite{mishra2004new,bu2003topological,sim2009mining}).

Bi-cliques are a fundamental concept in graph theory, and a variety of tasks on static and dynamic graphs are formulated as problems of finding (near) bi-cliques in them. Examples include:

\begin{itemize}[leftmargin=9pt]
    \item \textbf{Anomaly Detection~\cite{shin2017densealert,shin2018fast,shin2021detecting,jiang2015general}}: (Near) bi-cliques in real-world dynamic graphs signal anomalies, such as network intrusion~\cite{shin2018fast}, spam reviews~\cite{shin2021detecting}, edit wars on Wikipedia~\cite{shin2018fast}, and retweet boosting on Weibo~\cite{jiang2015general}.
    \item \textbf{Community Detection~\cite{kumar1999trawling,liu2006efficient,araujo2014com2}}: Relevant web pages often do not reference each other \cite{kumar1999trawling}, and thus finding (near) cliques can be ineffective for detecting them.
    Instead, related web pages can be detected by finding (near) bi-cliques formed between them and users visiting them~\cite{liu2006efficient}.
    Near bi-cliques were also searched for detecting temporal communities in phone-call networks \cite{araujo2014com2}.
    \item \textbf{Lossless Graph Compression~\cite{koutra2014vog, shah2015timecrunch, araujo2014com2}}: A static or dynamic graph $\GG$ can be described concisely but losslessly by (a) (near) bi-cliques and (b) the difference between $\GG$ and the graph described by the (near) bi-cliques~\cite{koutra2014vog,shah2015timecrunch, araujo2014com2}.
    \item \textbf{Phylogenetic Tree Construction~\cite{sanderson2003obtaining}}: A phylogenetic tree shows the evolutionary interrelationships among multiple species that are thought to have a common ancester.
    Identifying bi-cliques formed between genes and species containing the genes is helpful for accurate tree construction.
\end{itemize}
Additionally, (near) bi-cliques have been used for better understanding protein-protein interactions~\cite{bu2003topological} and stock markets~\cite{sim2009mining}. 

\begin{figure*}[t]
    \centering
    \vspace{-6mm}
    \subfloat[\small Speed and Quality]{
    \includegraphics[width=0.167\textwidth]{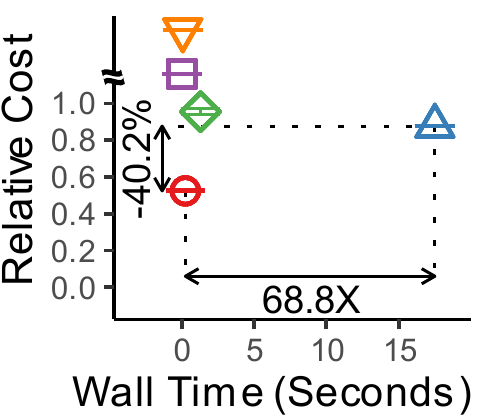}\label{subfig:tradeoff}
    \includegraphics[width= 0.112\textwidth]{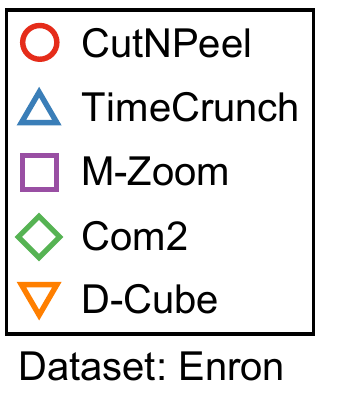}}
    \subfloat[\small Scalability]{\includegraphics[width=0.167\textwidth]{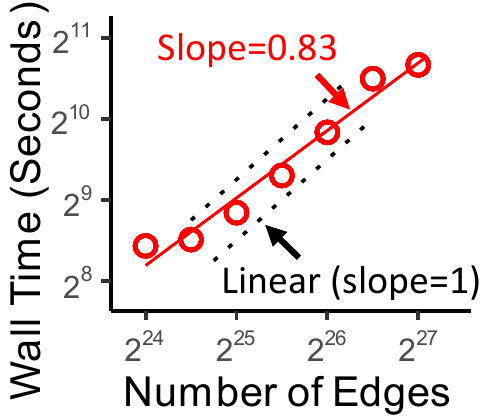}}
	\subfloat[\small Application: Pattern Discovery]{\includegraphics[width=0.36\textwidth]{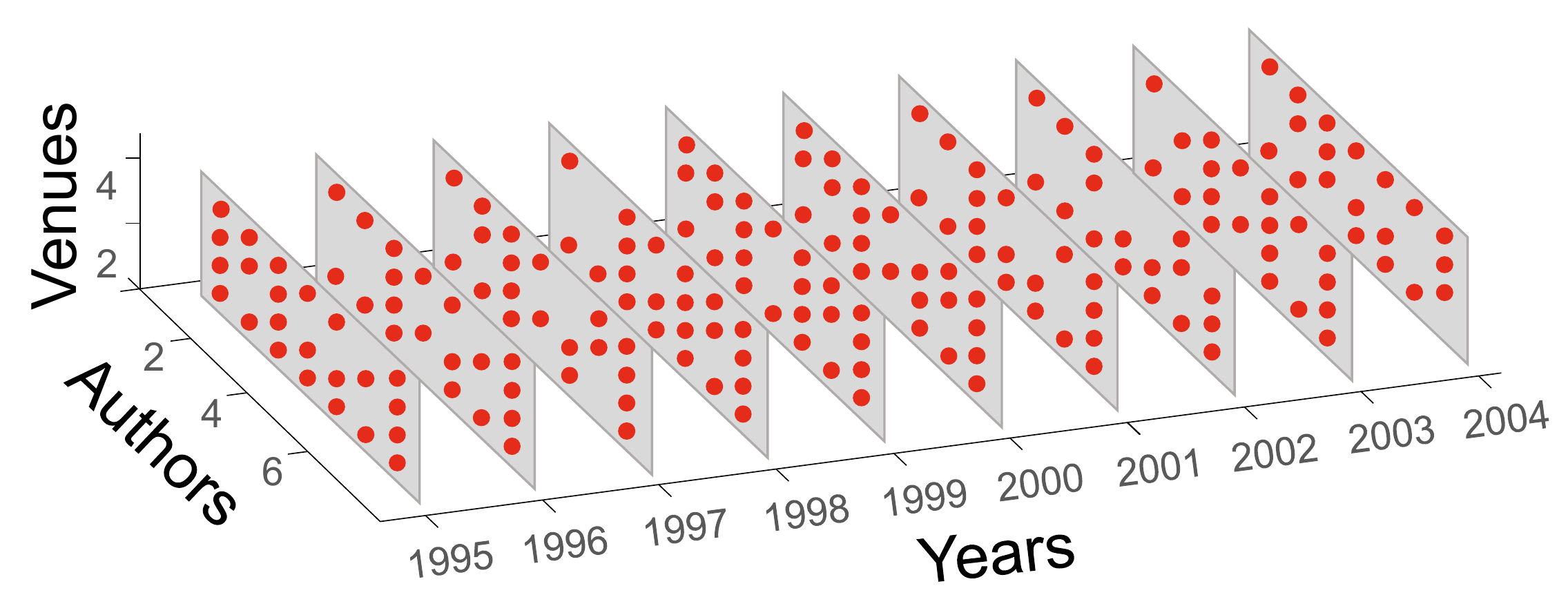}}
    \subfloat[\small Application: Compression]{
    \includegraphics[width=0.19\textwidth]{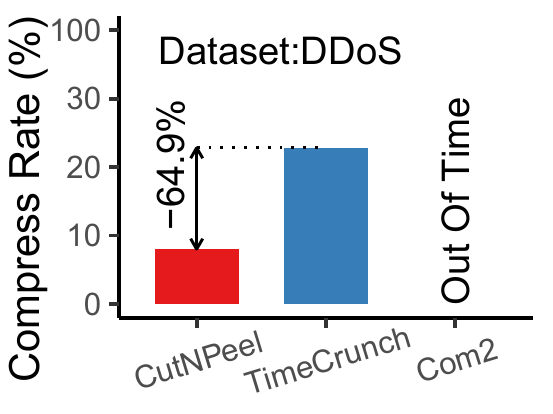}\label{subfig:cr}}\\
	\vspace{-1mm}
    \caption{\label{fig:highlights} Strengths of \method. See Section~\ref{sec:exp} for details. (a) \method finds high-quality near bi-cliques (i.e., those with low cost in Eq.~\eqref{eqn:obj}) rapidly. (b) \method scales to a graph with 134 million edges, near-linearly with the number of edges. (c) \method detects a long-lasting near bi-clique between four relevant venues (Int. Test Conf., Asian Test Symp., IEEE VLSI Test Symp., and VLSI Design Conf.) 
    and seven authors contributing to them. (d) \method achieves the best lossless compression.
    }
    
\end{figure*}

Due to the importance of (near) bi-cliques, a great number of search algorithms have been developed.
Most of them aim to enumerate all maximal (near) bi-cliques that maximize certain criteria within a static or dynamic graph~\cite{alexe2004consensus,makino2004new,liu2006efficient,kloster2019mining,dias2005generating,quine1955way,mishra2004new,bu2003topological,sim2009mining,das2018incremental}.
However, as the number of such subgraphs can be exponential in the number of nodes, finding all is practically impossible for large graphs and often unnecessary with many insignificant ones (e.g., too small or highly overlapped ones). Thus, several algorithms aim to find one or a predefined number of (near) bi-cliques that maximize certain criteria in a static or dynamic graph~\cite{shin2018fast,shin2021detecting, shin2017densealert, jiang2015general}.

This paper focuses on finding a ``good'' set of near bi-cliques since finding all (maximal) near bi-cliques is infeasible for large graphs and finding one or few of them is insufficient for many applications. Then, two questions naturally arise:
\begin{enumerate}[leftmargin=9pt]
    \item What is a ``good'' set of near bi-cliques within a graph? How should we measure the quality of a set of near bi-cliques?
    \item How can we rapidly detect a high-quality set of near bi-cliques, especially in large dynamic graphs?  
\end{enumerate}

Regarding the first question, we measure how \textit{concisely}, \textit{precisely}, and \textit{exhaustively} a given set of near bi-cliques describes the input dynamic graph $\GG$. That is, a high-quality set of near bi-cliques in $\GG$ consists of a \textit{tractable number} of subgraphs \textit{close to bi-cliques} that together \textit{cover a large portion of $\GG$}.
We take these three perspectives into account simultaneously in a systematic way based on the Minimum Description Length principle \cite{grunwald2007minimum,galbrun2020minimum}.

Regarding the second question, we propose \method (\textbf{Cut}ting a\textbf{N}d \textbf{Peel}ing), a fast search algorithm for a high-quality set of near bi-cliques in a dynamic graph.
As its name suggests, \method consists of the ``cutting'' step and the ``peeling'' step.
Specifically, it first partitions the input dynamic graph to reduce the search space, and then it finds near bi-cliques within each partition one by one by a top-down search. 
Thanks to our adaptive re-partitioning scheme, surprisingly, the cutting step not only reduces the search space but also improves the quality of near bi-cliques by guiding the  top-down search in the peeling step.

We demonstrate the effectiveness of \method, using six real-world dynamic graphs, though comparisons with four state-of-the-art algorithms \cite{araujo2014com2,shah2015timecrunch,shin2018fast,shin2021detecting} for finding near bi-cliques in dynamic graphs. In summary, \method has the following advantages:
\begin{itemize}[leftmargin=9pt]
    \item \textbf{High Quality}: \method provides near bi-cliques of up to $\mathit{51.2\%}$ \textit{better quality} than the second best method (Figure~\ref{fig:highlights}a). 
    \item \textbf{Speed}: \method is up to $\mathit{68.8\times}$ \textit{faster} than the competitors that is the second best in terms of quality (Figure~\ref{fig:highlights}a).
    \item \textbf{Scalability}: Empirically, \method scales near-linearly with the size of the input graph (Figure~\ref{fig:highlights}b).    
    \item \textbf{Applicability}: 
    Using \method, we achieve the best lossless compression and discover meaningful patterns (Figure~\ref{fig:highlights}c-d).
\end{itemize}
\textbf{Reproducibility}: The source code and datasets used in this paper can be found at \url{https://github.com/hyeonjeong1/cutnpeel}.

In Section~\ref{sec:concept}, we introduce some basic concepts.
In Section~\ref{sec:problem}, we formally define the problem of finding near bi-cliques in dynamic graphs.
In Section~\ref{sec:algo}, we present our proposed algorithm \method.
In Section~\ref{sec:exp}, we provide our experimental results.
In Section~\ref{sec:related}, we discuss related studies. 
In Section~\ref{sec:conclusion}, we conclude the paper.

\section{Basic Concepts}
    \label{sec:concept}
In this section, we introduce some basic concepts used throughout the paper. We list some frequently-used notations in Table~\ref{Tab:Tos}.

\begin{table}[t]
    \centering
    \small
	\caption{Table of frequently-used symbols \label{Tab:Tos}}
	\scalebox{0.86}{
	\begin{tabular}{r|l}
		\toprule
		\textbf{Symbol} & \textbf{Definition} \\
		\midrule
		$\Glong$ & input dynamic graph\\
		$\SSS,\D,\T$ & sets of source nodes, destination nodes, and timestamps in $\GG$\\
		$\I=\SSS\cup \D\cup\T$ & set of objects in $\GG$\\
	    $\E$ & set of edges in $\GG$ \\
		
		\midrule
		$\IT=\srcset\cup \dstset\cup\timeset$ & subset of the object set $\I$ where $\srcset\subseteq \SSS$,  $\dstset\subseteq \D$, and $\timeset \subseteq \T$
		\\
		$\GIT$ & near bi-clique composed of $\IT$\\
		$\EIT$ & set of edges in $\GIT$ \\
		$\PIT$ & exact bi-clique composed of $\IT$\\
		$|\PIT|$ & number of edges in $\PIT$ \\
		
		\midrule
		$\nearset$ & set of near bi-cliques\\
		$\objB$ & objective function to be minimized\\
		
		\midrule
		$\blockset,\correctionP,\correctionM$ & sets of exact bi-cliques, remaining edges, and missing edges  \\

		\midrule
		$\saving{\IT, \GP}$ & approximate saving in the objective due to  $\GIT$\\
		$\threshold{t}$ & threshold in each $t$-th iteration \\
		$\alpha,\iter$ & decrement rate of thresholds and the number of iterations \\
			$\Ggroups$ & set of partitions in $\GP$\\
		\bottomrule
	\end{tabular}
	}
\end{table}

\begin{figure*}[t]
    \vspace{-3mm} 
    \centering
    \includegraphics[width=\linewidth]{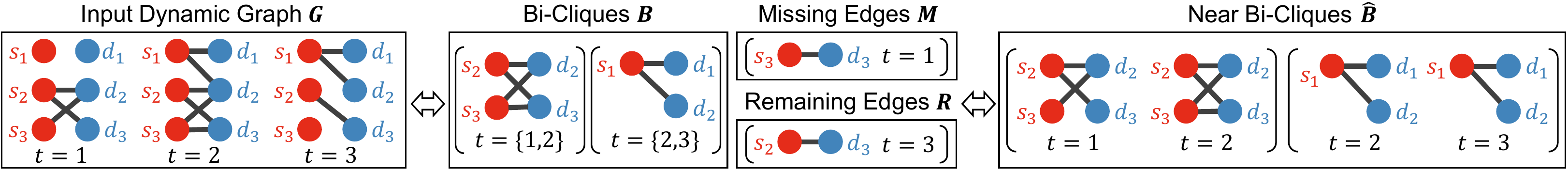} \\ \vspace{-1mm}
    \caption{The input dynamic graph $\Glong$ is described by exact bi-cliques $\blockset$, missing edges $\correctionM$, and remaining edges $\correctionP$.
    \label{fig:concept}}
\end{figure*}

\smallsection{Dynamic graphs}
A \textit{dynamic graph} 
$\Glong$ is a sequence of graphs over timestamps $\T$.
We denote $\GG$ at each time $t\in \T$ by $\GT$, where $\ST$ is the set of source nodes,  $\DT$ is the set of destination nodes, and $\ET$ is the set of edges.
We let $\SSS:=\bigcup_{t\in \T} \ST$ be the set of all source nodes and let $\D:=\bigcup_{t\in \T} \DT$ be the set of all destination nodes.
For simplicity, we assume $\SSS$, $\D$ and $\T$ are disjoint. That is, unipartite graphs are treated as bipartite graphs (i.e., a source and the same node as a destination are treated as different nodes). 
We use $\I:=\SSS \cup \D \cup \T$ to denote the set of \textit{objects} in $\GG$.
We use $\edge\in \SSS \times \D\times \T$ to indicate the edge $(\vi,\vj)\in \ET$ at time $t$, 
and we use $\E:=\{(s,d,t)\in \SSS \times \D\times \T :(s,d)\in \ET\}$ to indicate the set of all edges in $\GG$.
For any subset $\IT\subseteq \I$ of objects, we use $\GIT$ to denote the subgraph of $\GG$ induced by $\IT$ and use
$\EIT:=\{(s,d,t)\in \E: s \in \IT, d \in \IT, t \in \IT \}$ to denote the edges in it.


\smallsection{Bi-cliques in dynamic graphs}
A \textit{bi-clique} is a complete bipartite graph.
That is, a graph $\GTT$ with two disjoint node sets $\srcset$ and $\dstset$ is a bi-clique if every node in $\srcset$ is adjacent to every node $\dstset$, i.e., if $\edgeset = (\srcset \times \dstset)$.
A \textit{temporal bi-clique} is a bi-clique that appears once or repeatedly over time.
For any set $\IT$ of objects, we use $\PIT$ to denote the temporal bi-clique formed by the objects, and we use $|\PIT|$ to denote the number of edges in it.
That is, if $\IT=\srcset\cup \dstset\cup\timeset$ where $\srcset\subseteq \SSS$, $\dstset\subseteq \D$, and $\timeset \subseteq \T$, then $\PIT=(\GTTTshort)_{t\in \timeset}$ where $\GTTTshort=(\srcset \cup \dstset, \srcset \times \dstset)$ for all $t\in \timeset$, and $|\PIT|=|\srcset|\cdot|\dstset|\cdot|\timeset|$.

\smallsection{Near bi-cliques in dynamic graphs}
We use the term \textit{near bi-clique} to denote a bipartite graph ``close'' to a bi-clique with few or no of missing edges. Specifically, a subgraph $\GTT$ is a near bi-clique if it satisfies $\edgeset \approx (\srcset \times \dstset)$. 
Similarly, a dynamic graph $\GTlong$ with objects $\IT=\srcset \cup \dstset \cup \timeset$ and edges $\edgeset \subseteq \srcset \times \dstset \times \timeset$ is a \textit{near temporal bi-clique} if it satisfies $\edgeset \approx (\srcset \times \dstset \times \timeset)$.
Since this paper focuses on dynamic graphs, from now on, we will use the term \textbf{(near) bi-cliques} to refer to \textbf{(near) temporal bi-cliques} for ease of explanation.

\section{Problem Formulation}
\label{sec:problem}
In this section, we present how we measure the quality of a set of near bi-cliques. Then, we define the problem of finding a high-quality set of near bi-cliques in a dynamic graph.


\subsection{Quality of a Set of Near Bi-Cliques}
\label{sec:problem:quality}
In this subsection, we describe three sub-goals that any ``good'' set of near bi-cliques should pursue.
We also define cost functions that we aim to minimize to achieve the sub-goals.
The cost functions are commonly in bits so that we can directly compare and eventually balance them.
To this end, given a dynamic graph $\GG$ and a set $\nearset$ of near bi-cliques in it, we consider the following three sets:
\begin{itemize}[leftmargin=9pt]
    \item \textbf{Exact bi-cliques $\blockset$:} the set of minimal exact bi-cliques that contain the near bi-cliques in $\nearset$. That is, $\blockset:=\bigcup_{\GIT\in \nearset}\{\PIT\}$.
    \item \textbf{Missing edges $\correctionM$:} the set of \textit{missing edges}, which are included in  exact bi-cliques in $\blockset$ but not in $\GG$. 
    \item \textbf{Remaining edges $\correctionP$:} the set of \textit{remaining edges} in $\GG$ that do not belong to any near bi-clique in $\nearset$.
\end{itemize}
As illustrated in Figure~\ref{fig:concept}, the bi-cliques in $\blockset$ may contain missing edges, and
the near bi-cliques in $\nearset$ are obtained by simply removing the missing edges from each bi-clique in $\blockset$.



\smallsection{Sub-goal 1. (Preciseness)} The members of a ``good'' set of near bi-cliques should actually be close to exact bi-cliques. That is, the number of missing edges $\correctionM$ contained in the bi-cliques in $\blockset$ should be minimized. To pursue preciseness, we aim to minimize Eq.~\eqref{eq:preciseness}, which corresponds to the number of bits to encode $\correctionM$.
\begin{equation}
    \objF_{P}(\correctionM) := |\correctionM|\cdot\LE \label{eq:preciseness},
\end{equation}
where $\LE := \log_{2}|\SSS| + \log_{2}|\D| + \log_{2}|\T|$ corresponds to the number of bits to encode one missing edge.
We assume that the coordinate list (COO) format is used, and thus we list $\vi$, $\vj$ and $t$ to encode an edge $\edge$.
Additionally, we assume that $\log_{2}|\SSS|$ bits are required to encode a member of a set $\SSS$.

\smallsection{Sub-goal 2. (Exhaustiveness)} The members of a ``good'' set of near bi-cliques should cover a large portion of $\GG$. Equivalently, the number of remaining edges $\correctionP$ uncovered by any near bi-cliques in $\nearset$ should be minimized. To achieve exhaustiveness, we aim to minimize Eq.~\eqref{eq:exhaustiveness}, i.e., the number of bits to encode $\correctionP$.
\begin{equation}
    \objF_{E}(\correctionP) := |\correctionP|\cdot\LE. \label{eq:exhaustiveness}
\end{equation}

\smallsection{Sub-goal 3. (Conciseness)}
Note that Eq.~\eqref{eq:preciseness} and Eq.~\eqref{eq:exhaustiveness} can be trivially minimized by enumerating all edges, which are $1\times1$ bi-cliques, in $\GG$.
However, larger bi-cliques are more likely to be useful in applications mentioned in Section~\ref{sec:intro}, so larger ones should be preferred over trivial ones in a ``good'' set of near bi-cliques.
To design a cost function in bits, we note that larger bi-cliques can be used to encode the edges in them more concisely with fewer bits. Specifically, all edges in a bi-clique $\PIT$ where $\IT=\srcset \cup \dstset \cup \timeset$ can be encoded together by listing the objects that form it, requiring only $O(\IT)=O(|\srcset|+|\dstset|+|\timeset|)$ bits, instead of $O(|\srcset|\cdot |\dstset| \cdot |\timeset|)$ bits for encoding each  edge separately.
This saving in bits increases as bi-cliques become larger, as formalized in Lemma~\ref{lemma:conciseness}. 

\begin{lemma} \label{lemma:conciseness}
If a bi-clique $\PIT$ is strictly bigger\footnote{The number of object of each type in $\PIT$ is greater than or equal to that in $\PITP$, and the number of object of a type in $\PIT$ is strictly greater than that in $\PITP$.
} than a bi-clique $\PITP$, then
the number of bits per edge is smaller in $\PIT$ than in $\PITP$, i.e., \begin{equation}
    {\LPIT}\ / \ {|\PIT|} < {\LPITP}\ / \ {|\PITP|}. \label{eq:lemma:c}
\end{equation}
\begin{proof}
See Appendix~A \cite{onlinesupplementary}.
\end{proof}
\end{lemma}

Based on this observation, we aim to minimize Eq.~\eqref{eq:conciseness}, which is the number of bits to encode the edges contained in the bi-cliques in $\blockset$, to pursue conciseness. 
\begin{equation}
    \objF_{C}(\blockset) := \sum\nolimits_{\PIT\in \blockset} \LPIT, \label{eq:conciseness}
\end{equation}
where $\LPIT  :=(|\IT|+1)\cdot\log_{2} |\I|$ is the number of bits for encoding all edges in a bi-clique $\PIT$ together.\footnote{Additional $\log_{2} |\I|$ bits are used to encode the number of objects in $\PIT$.}
Note that in Eq.~\eqref{eq:conciseness}, edges that belong to multiple bi-cliques are encoded multiple times, increasing the cost. Thus, Eq.~\eqref{eq:conciseness} penalizes highly-overlapped bi-cliques, aligning with our pursuit of conciseness.

\smallsection{Total cost and the MDL principle}
Recall that the three cost functions (i.e., Eq.~\eqref{eq:preciseness}, Eq.~\eqref{eq:exhaustiveness}, and Eq.~\eqref{eq:conciseness}) are measured commonly in bits and thus directly comparable.
Since $\blockset$, $\correctionM$, and $\correctionP$ together exactly describe the input dynamic graph $\GG$, as shown in Figure~\ref{fig:concept}, minimizing their sum aligns with the Minimum Description Length (MDL) principle \cite{grunwald2007minimum}.
Thus, the \textit{total cost} for balancing our three sub-goals (i.e., preciseness, exhaustiveness, and conciseness) is defined as:
\begin{equation}
    \objB := \objF_{P}(\correctionM)+\objF_{E}(\correctionP)+ \objF_{C}(\blockset), \label{eqn:obj}
\end{equation}
where $\objF$ can be a function of $\nearset$ and $\GG$ since $\blockset$, $\correctionM$, and $\correctionP$ are directly obtained from $\nearset$ and $\GG$.



\subsection{Problem Definition}

Using the quality measure in Eq.~\eqref{eqn:obj}, we formalize our problem as:
\begin{problem}[Finding a Concise, Precise, and Exhaustive Set of Near Bi-Cliques in a Dynamic Graph]\label{defn:problem} \hfill \
\begin{itemize}[leftmargin=9pt]
    \item[] \textbf{(1) Given:} a dynamic graph $\dynamicGSum$,
    \item[] \textbf{(2) Find:} near bi-cliques $\nearset$
	\item[] \textbf{(3) To minimize:} $\objB$. 
\end{itemize}
\end{problem}




\section{Proposed Algorithm}
\label{sec:algo}
\begin{figure*}[t]
        \vspace{-3mm}
\end{figure*}

In this section, we propose \method, a fast algorithm for finding a high-quality set of near bi-cliques in dynamic graphs.
We first present its preliminary version \naive. Then, we present \method.

\subsection{\naive: Preliminary Algorithm} \label{sec:algo:melodyNaive}
In this subsection, we present \naive, a preliminary version of our proposed algorithm \method.
We provide an overview of \naive and then describe how it finds a single near bi-clique in detail.
After that, we analyze its complexity and lastly discuss its limitations.

\subsubsection{Overview (Algorithm~\ref{alg:overviewNaive})}
\naive repeatedly finds near bi-cliques one by one using a top-down search called \peeler (line~\ref{line:overviewNaive:peeler}).
Specifically, \peeler returns objects $\IT$, and the near bi-clique $\GIT$ formed by them in the original input graph $\GG$ is added to the output set $\nearset$ (line~\ref{line:overviewNaive:include}).
Whenever \naive finds a near bi-clique, the edges in it are removed from the input graph to prevent \naive from finding the same near bi-clique (line \ref{line:overviewNaive:update}), and we use $\GP$ to denote the dynamic graph with the remaining edges.
\naive stops searching when $\saving{\IT,\GP}$ in Eq.~\eqref{eqn:saving} is zero or negative for found objects $\IT$ (lines~\ref{line:overviewNaive:terminate1}-\ref{line:overviewNaive:terminate2}) and returns $\nearset$ as the final output (line~\ref{line:overviewNaive:return}).
\begin{equation}
\label{eqn:saving}
	\saving{\IT,\GP} := \objF(\emptyset,\GP)-\objF(\{\GPIT\},\GP),
\end{equation}
where $\objF(\emptyset,\GP)$ is the total cost $\objB$ when $\nearset=\emptyset$ and $\GG=\GP$; and  $\objF(\GIT,\GP)$ is that when $\nearset=\{\GPIT\}$ and $\GG=\GP$.
We use $\saving{\IT,\GP}$ in Eq.~\eqref{eqn:saving} to approximate $\objF(\nearset,\GG)-\objF(\nearset\cup\{\GIT\},\GG)$, i.e., the amount of saving in the total cost due to a new near bi-clique $\GIT$.
$\saving{\IT,\GP}$ can be computed in $O(1)$ time, as shown in Lemma~\ref{lemma:approximation}, while providing a preciseness guarantee in Theorem~\ref{thm:preciseness}. 


\begin{lemma} \label{lemma:approximation}
Given the numbers of objects of each type and edges in $\GP$ and $\GPIT$,  $\saving{\IT,\GP}$ can be computed in $O(1)$ time.
\begin{proof}
If $\IT=\srcset\cup \dstset \cup  \timeset$ where $\srcset\subseteq \SSS$, $\dstset\subseteq \D$, and $\timeset\subseteq \T$, then Eqs.~\eqref{eq:preciseness}-\eqref{eqn:saving} imply Eq.~\eqref{eq:approximiation}, where every term is given.
\begin{equation}
    \saving{\IT,\GP} =  
    (2\cdot |\EPIT| - |\srcset|\cdot |\dstset| \cdot |\timeset|)\cdot\LE - (|\IT|+1)\cdot\log_{2} |\I|. \label{eq:approximiation} \qedhere
\end{equation}
\end{proof}
\end{lemma}

\begin{thm}[Preciseness of \naive] \label{thm:preciseness}
The density of every near bi-clique $\GIT\in \nearset$ obtained by Algorithm~\ref{alg:overviewNaive} is greater than $0.5$, i.e., 
\begin{equation}
    {|\EIT|}\ / \ {|\PIT|}>0.5. \label{eq:guarantee:preciseness}
\end{equation}
\begin{proof}
By lines \ref{line:overviewNaive:terminate1} and \ref{line:overviewNaive:terminate2} of Algorithm~\ref{alg:overviewNaive}, Eq.~\eqref{eq:positive} holds.
\begin{equation}
    \saving{\IT,\GP}>0, \ \forall \GIT\in \nearset. \label{eq:positive}
\end{equation}
Eq.~\eqref{eq:approximiation}, Eq.~\eqref{eq:positive}, and $|\EIT|\geq |\EPIT|$ imply Eq.~\eqref{eq:guarantee:preciseness}.
\end{proof}
\end{thm}

\subsubsection{Finding one near bi-clique (Algorithm~\ref{alg:peeler})}
We describe \peeler, which is a subroutine of \naive for finding a single near bi-clique. 
Starting from the set $\I$ of all objects in $\GP$ (line~\ref{line:peeler:init}), \peeler repeatedly removes an object  (line~\ref{line:peeler:remove}) until an empty set is left.
When choosing the object to be removed, \peeler chooses one with the sparsest connectivity for a near bi-clique to remain. 
Specifically, it chooses an object $i\in \I$ that minimizes $\density{\IT, i}$ in Eq.~\eqref{eqn:density} (line~\ref{line:peeler:ifRemoved}).
\begin{equation}
\label{eqn:density}
    \density{\IT, i} := \frac{|\EIT|-|\EITM|}{|\PIT|-|\PITM|},
\end{equation}
where $|\PIT|-|\PITM|$ is the number of potential edges adjacent to $i$, and $|\EIT|-|\EITM|$ is the number of existing edges among them.
As the set $\IT$ of remaining objects changes, \method tracks $\saving{\IT,\GP}$ in Eq.~\eqref{eqn:saving} (lines~\ref{line:peeler:update1}-\ref{line:peeler:update2}).
Lastly, as its final output, \peeler returns $\IT$ when $\saving{\IT,\GP}$ is maximized (lines~\ref{line:peeler:output1}-\ref{line:peeler:output2}).   

\begin{algorithm}[t]
    \small
	\caption{\label{alg:overviewNaive} Overview of \naive}
	\KwIn{dynamic graph $\GG$ with objects $\I$ and edges $\E$
	}
	\KwOut{near bi-cliques $\nearset$}
	initialize $\GP$ with objects $\I$ and edges $\EP=\E$\\
	\While{true}{
	    $\IT \leftarrow \peeler(\GP)$\label{line:overviewNaive:peeler}  \Comment{Algorithm~\ref{alg:peeler}}\\
	    \If{$\saving{\IT,\GP} \leq 0$ \label{line:overviewNaive:terminate1}}{
    	    \textbf{break}\label{line:overviewNaive:terminate2}
	    }
	    $\nearset \leftarrow \nearset \cup \{\GIT\}$\label{line:overviewNaive:include}\\
	    $\EP \leftarrow \EP-\EPIT$\label{line:overviewNaive:update} \\
	}
	\Return $\nearset$ \label{line:overviewNaive:return} 
\end{algorithm}
\begin{algorithm}[t] 
    \small
	\caption{\label{alg:peeler} \peeler: top-down search of a bi-clique}
	\KwIn{dynamic graph $\GP$ with objects $\I$ and edges $\EP$
	}
	\KwOut{objects $\IT_{max}$ that form a near bi-clique }
	$\IT \leftarrow \I$; \ \  $\densityMax \leftarrow - \infty$; \ \  $t_{max} \leftarrow 0$ \label{line:peeler:init}\\
	\For{$t = 1, \cdots, |\I|$\label{line:peeler:while}}
	{
	    \If{$\saving{\IT,\GP}>\densityMax$}{\label{line:peeler:update1}
	        $\densityMax\leftarrow \saving{\IT,\GP}$; \ \  $t_{max}\leftarrow t$ \label{line:peeler:update2}\\
	    }
	    $i_{t} \leftarrow i \in \IT$ with minimum $\density{\IT, i}$ \label{line:peeler:ifRemoved}\\
	    $\IT  \leftarrow \IT \setminus \{i_t\}$ \label{line:peeler:remove}\\
	    
	}
	$\IT_{max}\leftarrow \{i_{t_{max}},i_{t_{max}+1},\cdots,i_{|\I|}\}$ \label{line:peeler:output1} \\
	\Return $\IT_{max}$ \label{line:peeler:output2}
\end{algorithm}

\subsubsection{Relation with \cite{shin2018fast}}
\naive is based on the top-down greedy search framework \cite{shin2018fast}, which is originally designed for detecting dense tensors (i.e., multi-dimensional arrays). 
\naive uses the novel selection functions in Eq.~\eqref{eqn:saving} and Eq.~\eqref{eqn:density}, instead of those suggested in \cite{shin2018fast},\footnote{For example, $|\EPIT|/|\IT|$ and $|\EPIT|/(|\PIT|)^{1/3}$.} 
and this change significantly improves the quality of detected near bi-cliques, as shown empirically in Section~\ref{sec:exp:quality} and Appendix~D \cite{onlinesupplementary} (compare the relative cost of M-Zoom \cite{shin2018fast} in Figure~\ref{fig:quality} and that of \naive in Figure~6a in \cite{onlinesupplementary}).
Specifically, the selection functions suggested in \cite{shin2018fast} do not guarantee Eq.~\eqref{eq:guarantee:preciseness}, and empirically, using them gives near bi-cliques that lack preciseness (see Figure~\ref{fig:patterns} in Section~\ref{sec:exp:quality}).





\subsubsection{Complexity}
We present the time and space complexities of \naive in Theorems~\ref{thm:naive:complexity:time} and
\ref{thm:naive:complexity:space}. For simplicity, we assume $|\I|=O(|\E|)$.

\begin{thm}[Time Complexity of \naive] \label{thm:naive:complexity:time}
The time complexity of Algorithm~\ref{alg:overviewNaive} is $O(|\E|^2 \cdot \log |\I|)$.
\begin{proof}
    Eq.~\eqref{eq:approximiation} and Eq.~\eqref{eq:positive} imply $|\EPIT|\geq 1$ for all $\GIT\in \nearset$. This and $\sum_{\GIT\in \nearset}|\EPIT|\leq |\E|$ imply $|\nearset|\leq |\E|$. That is, 
    the number of detected near bi-cliques is at most $|\E|$.
    From Lemma~\ref{lemma:approximation} and the aforementioned connection with \cite{shin2018fast}, it can be shown that finding one near bi-clique takes $O(|\E|\cdot \log |\I|)$ time (see Theorem~1 of \cite{shin2018fast}).
    Thus, the total time complexity is $O(|\E|^2 \cdot \log |\I|)$.
\end{proof}
\end{thm}

\begin{figure*}[t]
        \vspace{-3mm}
\end{figure*}

\begin{thm}[Space Complexity of \naive] \label{thm:naive:complexity:space}
The space complexity of Algorithm~\ref{alg:overviewNaive} is $O(|\E| \cdot \log |\I|)$.
\begin{proof}
   Eq.~\eqref{eq:approximiation} and Eq.~\eqref{eq:positive} imply Eq.~\eqref{eq:space}.
   \begin{equation}
       |\EPIT| \cdot\LE > (|\PIT| - |\EPIT|)\cdot\LE + (|\IT|+1)\cdot\log_{2} |\I|,  \ \forall \GIT\in \nearset, \label{eq:space}
   \end{equation} 
   where the right hand side upper bounds the number of bits required for storing $\GIT$.
   Eq.\eqref{eq:space} and $\sum_{\GIT\in \nearset}|\EPIT|\leq |\E|$ imply that storing the output $\nearset$ requires $O(|\E|\cdot\LE)=O(|\E| \cdot \log |\I|)$ bits.
   Additionally, storing $\GG$ and $\GP$ requires $O(|\E|\cdot\LE)=O(|\E| \cdot \log |\I|)$ bits, and storing  $\{i_t\}_{t=1}^{|\I|}$ in Algorithm~\ref{alg:peeler} requires $O(|\I| \cdot \log |\I|)$ bits.
   Thus, the total space complexity is $O(|\E| \cdot \log |\I|)$.
\end{proof}

\end{thm}




\subsubsection{Limitations}
Finding a large number of near bi-cliques using \naive suffers from high computational cost since \naive needs to access all remaining edges (i.e., $\EP$) to detect each near bi-clique.
Moreover, \naive often fails to detect bi-cliques formed by objects whose connectivity outside the bi-cliques are sparse.
This is because such objects are likely to be removed in the early stage of \peeler, which prioritizes objects with dense connectivity.

\subsection{\method: Proposed Algorithm} \label{sec:algo:melodyAlgo}

\begin{algorithm}[t]
    \small
	\caption{\label{alg:overview} Overview of \method}
	\KwIn{
		\begin{enumerate}[leftmargin=9pt]
			\item dynamic graph $\GG$ with objects $\I=\SSS\cup\D\cup \T$ and edges $\E$
			\item maximum number of iterations $\iter$ ($\geq 1$)
			\item threshold decrement rate $\decrement$ ($\in (0,1)$)
		\end{enumerate}
	}
	\KwOut{near bi-cliques $\nearset$}
	
	initialize $\GP$ with objects $\I$ and edges $\EP=\E$ \label{line:overview:init} \\
	$\threshold{1}\leftarrow \infty$\\
	\For{$t = 1, \cdots, \iter$}
	{
	    $\threshold{t+1}\leftarrow 0$\\
		\For{\normalfont{\textbf{each}} object set $\V\in\{\SSS, \D, \T\}$ \label{line:overview:order}} 
		{   
		    $\Ggroups \leftarrow \dividing(\GP, \V)$ \label{line:overview:division}  \Comment{Algorithm~\ref{alg:dividing}}\\
		    \For{\normalfont{\textbf{each}} partition $\Gi{k}=(\Vi,\Ei)\in \Ggroups$}{
        	    \While{true}{
            	    $\IT \leftarrow \peeler(\Gi{k})$\label{line:overview:detect}  \Comment{Algorithm~\ref{alg:peeler}}\\
            	    \eIf{$\saving{\IT,\Gi{k}} < \threshold{t}$\label{line:overview:fails}}{
	                $\threshold{t+1} \leftarrow \max(\saving{\IT,\Gi{k}}\cdot\alpha, \threshold{t+1})$\label{line:overview:update}\\
        	        \textbf{break} the while loop\label{line:overview:faile}
	                }{
	                $\nearset \leftarrow \nearset \cup \{\GIT\}$\label{line:overview:include}\\ \label{line:overview:add}
	                $\Ei \leftarrow \Ei-\EPIT$\label{line:overview:exclude}\\ \label{line:overview:update2}
            	    $\EP \leftarrow \EP-\EPIT$\label{line:overview:update} \\ \label{line:overview:update3}
	                }
	            }
	        }
		}
	    \If{$\threshold{t+1} = 0$ \label{line:overview:terminate:conditions1}}{\textbf{break} \label{line:overview:terminate}}
	}
	\Return $\nearset$ \label{line:overview:terminate}
\end{algorithm}

In this subsection, we present \method, our proposed algorithm for Problem~\ref{defn:problem}.
We first discuss the main ideas behind \method. Then, we present an outline of \method.
After that, we describe its components in detail. Lastly, we compare it with \naive.

\subsubsection{Main ideas}
\label{sec:algo:melodyAlgo:idea}
In order to address the two aforementioned limitations of \naive, \method first partitions the input dynamic graph and then finds near bi-cliques within each partition.
Partitioning reduces the search space, and as a result, it reduces the computational cost for finding each near bi-clique.
Partitioning is also helpful to detect bi-cliques formed by objects with sparse global connectivity (i.e., sparse connectivity outside the bi-cliques).
This is because \method prioritizes objects based on their connectivity within partitions, instead of their global connectivity.
By reducing the search space, partitioning may also have a harmful impact on search accuracy. 
In order to minimize such a harmful impact, 
\method uses adaptive re-partitioning while employing randomness.

\subsubsection{Overview (Algorithm~\ref{alg:overview})}
Given an input dynamic graph $\GG$, the maximum number of iteration $\iter$, and the threshold decrement rate $\decrement$, \method returns near bi-cliques $\nearset$ in $\GG$.
As in \naive, \method starts a search from $\GG$ (line~\ref{line:overview:init}), and as it detects near bi-cliques, removes the edges belonging to the near bi-cliques from $\GP$ (lines~\ref{line:overview:update2} and \ref{line:overview:update3}), where $\GP$ denotes the graph with remaining edges.
At each iteration, \method partitions $\GP$ into subgraphs based on an object set $\V \in \{\SSS, \D, \T\}$ using \dividing (line~\ref{line:overview:division}), which is described in detail later. \dividing employs randomness so that different partitions are obtained at different iterations.
Within each partition $\Gi{k}$, \method repeatedly detects near bi-cliques using \peeler (line~\ref{line:overview:detect}) as in \naive.
Since the current partitions may not be ideal, \peeler stops finding near bi-cliques within a partition if $\saving{\IT,\Gi{k}}$ in Eq.~\eqref{eqn:saving} is less than a threshold $\threshold{t}$ for a found near bi-clique $\GIT$ (lines~\ref{line:overview:fails}-\ref{line:overview:faile}).
\method terminates (line~\ref{line:overview:terminate}) if $\threshold{t}$ reaches 0 (line~\ref{line:overview:terminate:conditions1}) or the number of iterations reaches $T$.

\subsubsection{Adaptive thresholding}
The threshold $\threshold{t}$ decreases adaptively as iterations proceed. Specifically, the threshold $\threshold{t+1}$ at the $(t+1)$-th iteration is set to $\alpha\in(0,1)$ of the maximum $\saving{\IT,\Gi{k}}$ value, among those less than $\threshold{t}$, at the $t$-th iteration.
The threshold balances exploration and exploitation. Specifically, in early iterations, \method puts more emphasis on 
exploration (i.e., exploring near bi-cliques in partitions in later iterations) by accepting near bi-cliques under strict conditions (i.e., with large $\threshold{t}$). In later iterations, it puts more emphasis on exploitation (i.e., finding near bi-cliques in current partitions) by accepting near bi-cliques under lenient conditions (i.e., with small $\threshold{t}$).


\begin{algorithm}[t]
    \small
	\caption{\label{alg:dividing} \dividing : partitioning $\GG$ into subgraphs}
	\KwIn{
	    \begin{enumerate}[leftmargin=9pt]
			\item dynamic graph $\GP$ with objects $\I=\SSS\cup \D \cup \T$ and edges $\EP$
			\item given a set of objects $\V \in \{\SSS, \D, \T\}$
		\end{enumerate}
	}
	\KwOut{disjoint subgraphs of graph, $\Ggroups$}
	Generate a uniform random $1$-to-$1$ function $h : \I \rightarrow \{1, \cdots, |\I|\}$\\ \label{line:dividing:hash}
	\For{\normalfont{\textbf{each}} object $\vvv \in \V$ \label{line:dividing:shingle:start} }
	{
	    $\shingleGraph{\vvv}\leftarrow \infty$ \\
	    \For{\normalfont{\textbf{each}} edge $e \in  \EP_{\I}-\EP_{\I\setminus \{v\}}$}{
    	$\shingleGraph{\vvv} \leftarrow \min(g_v(e),f(v))$ \label{line:dividing:shingle} \Comment{see Eq.~\eqref{eq:gfunction} for $g_v(e)$} \label{line:dividing:shingle:end}\\
    	}
	}
	partition $\V$ into $\Vgroups$ based on $\shingleGraph{\cdot}$ values \label{line:dividing:group}\\
	\For{each $\Vk \in \Vgroups$}{\label{line:dividing:partition1}
	    $\Gi{k}\leftarrow (\Vi, \EP_{\Vi})$
	}\label{line:dividing:partition2}
	\Return $\Ggroups$
\end{algorithm}


\begin{figure*}[t]
        \vspace{-7mm}
\end{figure*}

\subsubsection{\dividing for partitioning (Algorithm~\ref{alg:dividing})}
Below, we propose \dividing, a dynamic-graph partitioning algorithm used by \method.
When designing \dividing, we pursue the following goals: 
\begin{itemize}[leftmargin=9pt]
    \item{\textbf{G1.}} Speed: \dividing should be fast since it is performed repeatedly.
    \item{\textbf{G2.}} Locality: \dividing should locate each near bi-clique inside a partition, to make it detectable, rather than splitting it into partitions.
    \item{\textbf{G3.}} Randomness: \dividing should employ randomness so that more possibilities can be explored by repeatedly performing it.
\end{itemize}
\dividing partitions a given dynamic graph $\GP$ based on a given object set $\V \in \{\SSS, \D, \T\}$.
It first generates a uniform random $1$-to-$1$ function $h : \I \rightarrow \{1, \cdots, |\I|\}$ for the set $\I$ of all objects (line~\ref{line:dividing:hash}), bringing randomness for \textbf{G3}.
Using the function $h$, we compute a function $f: \V \rightarrow \mathbb{Z}_{>0}$ (lines~\ref{line:dividing:shingle:start}-\ref{line:dividing:shingle:end}), based on which $\V$ is partitioned (line~\ref{line:dividing:group}).
Specifically, for each object $v\in \V$, $f(v)$ is defined as: 
\begin{equation*}
f(v):=\min\{g_v(e)\}_{e\in \EP_{\I}-\EP_{\I\setminus \{v\}}},
\end{equation*}
where $\EP_{\I}-\EP_{\I\setminus \{v\}}$ is the set of edges adjacent to $v$ and
\begin{equation}
    \label{eq:gfunction}
    g_v(e=(s,d,t)) := \begin{cases}
                        h(d)\cdot |\I|+h(t) & \text{ if } v=s \\
                        h(t)\cdot |\I|+h(s) & \text{ if } v=d \\
                        h(s)\cdot |\I|+h(d) & \text{ if } v=t.
                    \end{cases}
\end{equation}
Our design of $f(\cdot)$ is inspired by min-hashing~\cite{broder2000min}, and two objects in $\V$ are more likely to have the same $f(\cdot)$ value as their connectivity to other objects are more similar.
That is, \dividing pursues \textbf{G2} by making objects with similar connectivity be likely to belong to the same partition.
Lastly, edges in $\GP$ are partitioned according to the partitions of their incident objects in $\V$ (lines~\ref{line:dividing:partition1} and \ref{line:dividing:partition2}).
Regarding \textbf{G1}, \dividing takes linear time as formalized in Theorem~\ref{thm:cut:complexity:time}.

\begin{thm}[Time Complexity of \dividing] \label{thm:cut:complexity:time}
The time complexity of Algorithm~\ref{alg:dividing} is $O(|\I|+|\EP|)$.
\begin{proof}
    Generating $h(i)$ for all $i\in \I$ takes $O(|\I|)$ time. Computing $f(v)$ for all $v\in \V$ takes $O(|\V|+|\EP|)=O(|\I|+|\EP|)$ time. Partitioning $\V$ takes $O(|\V|)=O(|\I|)$ time, and partitioning $\E$ takes $=O(|\EP|)$ time. Hence, it takes $O(|\I|+|\EP|)$ time in total.
\end{proof}
\end{thm}

\smallsection{Comparison with \naivenosep}
Below, we assume $|\I|=O(|\E|)$ for ease of explanation.
For each near bi-clique, \method needs to access only the remaining edges within a partition (e.g., $\Ei$), while \naive needs to access all remaining edges (i.e., $\EP$). However, a partition can be as large as the entire dynamic graph.
Thus, the worst-case time complexity of \method is $O(|\E|^2 \cdot \log |\I|)$, as in \naive, if we assume $T=O(|\E|\cdot \log |\I|)$ so that performing \dividing $T$ times takes $O(T(|\I|+|\EP|))=O(|\E|^2 \cdot \log |\I|)$ time.
Compared to \naive, \method additionally stores $h(i)$ for all $i\in \I$ and $f(v)$ for all $v\in \V$, which take only $O(|\I|)=O(|\E|)$ space. Regarding preciseness, Eq.~\eqref{eq:guarantee:preciseness} still holds for every near bi-clique $\GIT\in \nearset$ from \method.

Empirically, however, \method scales near linearly with the size of the input dynamic graph, as shown experimentally in Section~\ref{sec:exp:scalability}. Moreover, as shown in Appendix~D \cite{onlinesupplementary}, \method
detects better near bi-cliques faster than \naive.


\section{Experiments}
\label{sec:exp}
\begin{table}[t]
    \centering
	\caption{Summary of the six real-world dynamic graphs. \label{Tab:data}} 
	\scalebox{0.75}{
	\begin{tabular}{c|c|c|c}
		\toprule
		\textbf{Name} & \textbf{Description} & \textbf{\# of Objects} & \textbf{\# of Edges}\\
		\midrule
	    Enron~\cite{shetty2004enron} & sender / receiver / time [week] & $140$ / $144$ / $128$ & $11,568$\\
		Darpa~\cite{lippmann2000evaluating} & Src IP / Dst IP / time [day] & $9,484$ / $23,398$ / $57$ & $140,069$\\
		DDoS~\cite{ddosdataset} & Src IP / Dst IP / time [second] & $9,312$ / $9,326$ / $3,954$ & $22,844,324$\\
        DBLP~\cite{dblpdataset} & author / venue / time [year] & $418,236$ / $3,566$ / $49$ & $1,325,416$\\
		Yelp~\cite{yelpdataset} & user / business / time [month] & $552,339$ / $77,079$ / $134$ & $2,214,201$\\
		Weeplaces~\cite{weeplacedataset} & user / place / time [month] & $15,793$ / $971,308$ / $92$ & $3,970,922$\\
		\bottomrule
	\end{tabular}
	}
\end{table}

In this section, we review our experiments to answer Q1-Q3.
\begin{itemize}[leftmargin=9pt]
    \item{\textbf{Q1. Search Quality \& Speed:}} How rapid is \method? Are identified near bi-cliques precise, exhaustive, and concise?
    \item{\textbf{Q2. Scalability:}} How does the running time of \method scale with respect to the size of the input graph?
    \item{\textbf{Q3. Application 1 - Compression:}} How concisely can we represent dynamic graphs using the outputs of $\method$?
\end{itemize}

\begin{figure*}[t]
    \vspace{-3mm}
	\centering
	\includegraphics[width=0.42\linewidth]{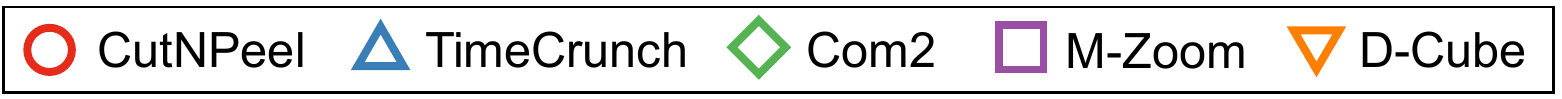} \\
	\vspace{-3mm}
    \subfloat[\small  Enron]{\includegraphics[width=0.167\textwidth]{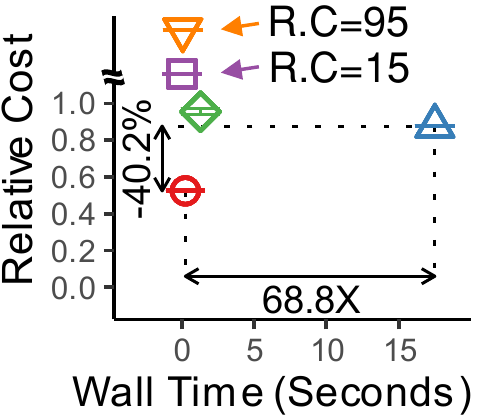}\label{subfig:q1enron}}
    \subfloat[\small  Darpa]{\includegraphics[width=0.167\textwidth]{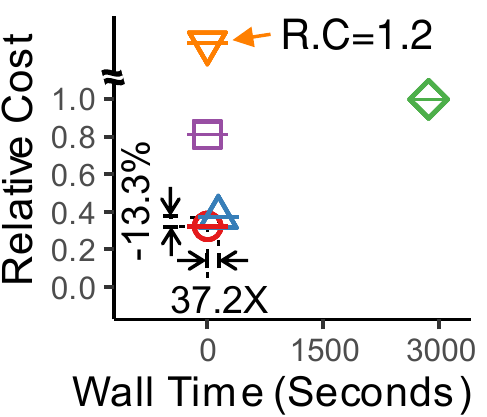}\label{subfig:q1darpa}}
    \subfloat[\small  DDoS]{\includegraphics[width=0.167\textwidth]{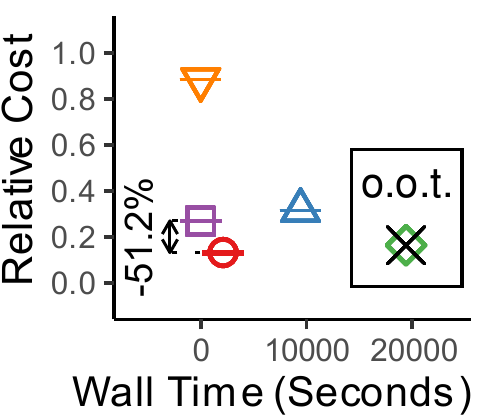}\label{subfig:q1ddos}} 
    \subfloat[\small  DBLP]{\includegraphics[width=0.167\textwidth]{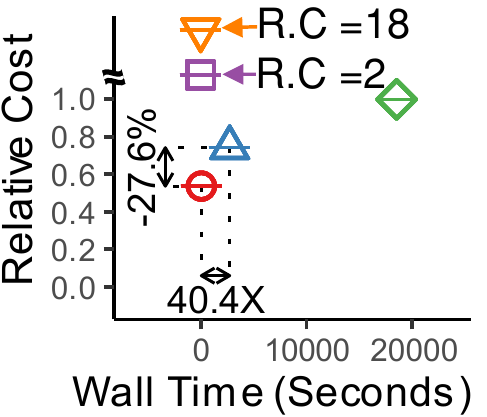}\label{subfig:q1dblp}}
    \subfloat[\small  Yelp]{\includegraphics[width=0.167\textwidth]{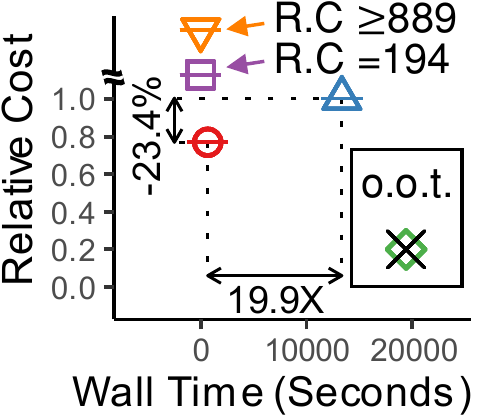}\label{subfig:q1Yelp}} 
    \subfloat[\small  Weeplaces]{\includegraphics[width=0.167\textwidth]{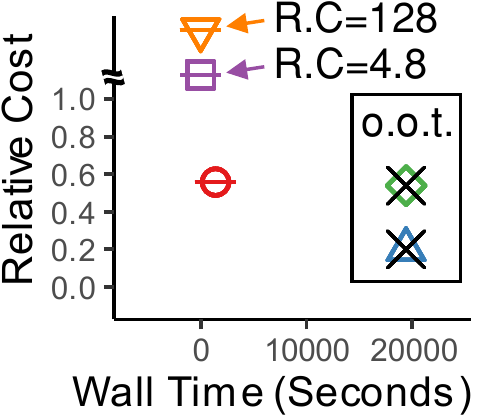}\label{subfig:q1weeplace}} \\
    \vspace{-1mm}
    \caption{\label{fig:quality}\textbf{\method is fast while providing high-quality near bi-cliques.} o.o.t.: out of time ($> 6$ hours). R.C: relative cost in Eq.~\eqref{eqn:cr} (the lower, the better). \method yields near bi-cliques with up to $\mathbf{51.2\%}$ \textbf{better quality}, up to $\mathbf{68.8\times}$ \textbf{faster}, than the competitors with the second best quality.
    We report the means of five trials, and error bars indicate $\pm 1$ standard deviation.
    }
\end{figure*}

\begin{figure*}[t]
    \centering
    \vspace{-2mm}
	\includegraphics[width= 0.42\linewidth]{FIG/legend.pdf} \\
	\vspace{-4mm}
    \subfloat[\small  Enron (\#=$979$)]{\includegraphics[width=0.167\textwidth]{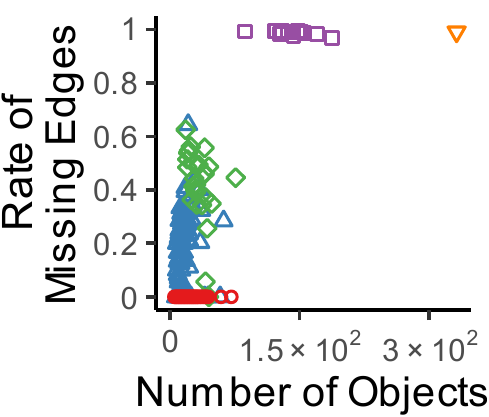}\label{subfig:patternenron}}
    \subfloat[\small  Darpa (\#=$2,981$)]{\includegraphics[width=0.167\textwidth]{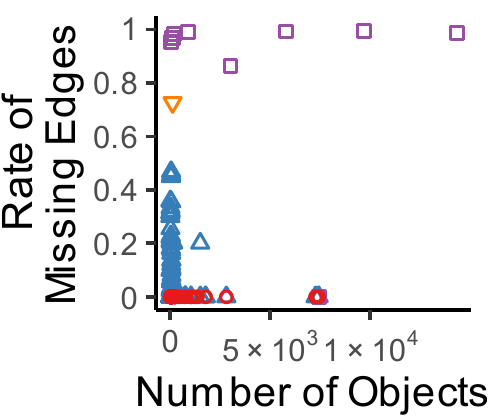}\label{subfig:patterndarpa}}
    \subfloat[\small  DDoS (\#=$7,840$)]{\includegraphics[width=0.167\textwidth]{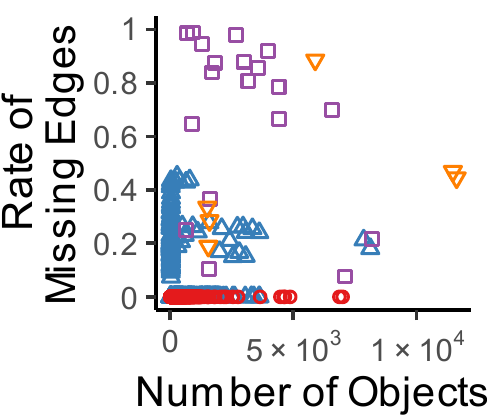}\label{subfig:patternddos}} 
    \subfloat[\small  DBLP (\#=$24,753$)]{\includegraphics[width=0.167\textwidth]{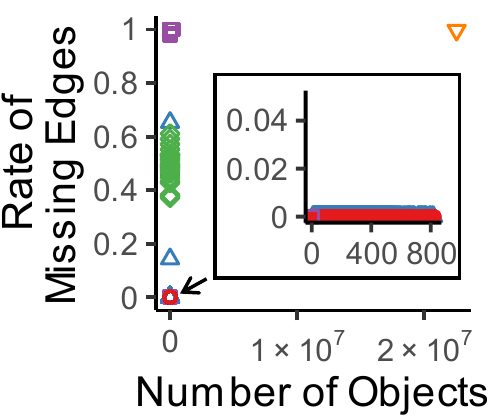}\label{subfig:patterndblp}}
    \subfloat[\small  Yelp (\#=$225,545$)]{\includegraphics[width=0.167\textwidth]{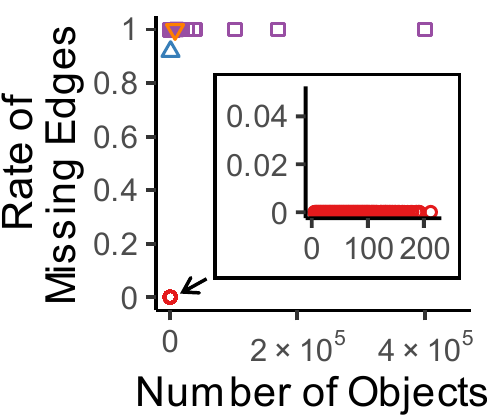}\label{subfig:patternYelp}} 
    \subfloat[\small Weeplaces (\#=$198,747$)]{\includegraphics[width=0.167\textwidth]{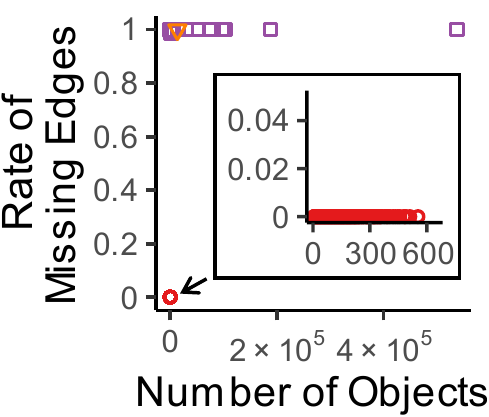}\label{subfig:patternweeplace}} \\
    \vspace{-1mm}
    \caption{\label{fig:patterns} \textbf{\method detects subgraphs close to bi-cliques.} Each plot shows the size and preciseness of the near bi-cliques detected by the considered algorithms in each dataset. 
    In parentheses, the number of near bi-cliques detected by \method in each dataset is reported.
    The near bi-cliques detected by \method tend to be precise with smaller ratios of missing edges. The number of near bi-cliques detected by \method is tractable and  much smaller than the number of edges. 
    } 
\end{figure*}

\subsection{Experiment Specifications}

\smallsection{Machines}
We used a machine with a 3.8GHz AMD Ryzen 3900X CPU and 128GB RAM for the scalability test in Section~\ref{sec:exp:scalability} and a machine with 3.7GHz i9-10900K CPU and 64GB RAM for the others.

\smallsection{Datasets}
We used six real-world dynamic graph datasets that are briefly described in Table~\ref{Tab:data}. 

\smallsection{Competitors and parameters}
We implemented \method in Java and set $\iter=80$ and $\alpha=0.8$ unless otherwise stated.
We compared \method with four competitors: \timecrunch \cite{shah2015timecrunch}, \comtwo \cite{araujo2014com2}, \mzoom \cite{shin2018fast}, and \dcube \cite{shin2021detecting}.
For the competitors, we used the official implementations provided by the authors, which are in Java (\comtwo, \mzoom, and \dcube) or MATLAB (\timecrunch).
For \timecrunch, we used the step-wise heuristic, which performed best in \cite{shah2015timecrunch}, and excluded chain subgraphs, which cannot be considered as bi-cliques, from its outputs.
For \dcube, we set $\theta=1$, which performed best in \cite{shin2021detecting}.
For \mzoom, and \dcube, we used the density function that minimized the cost function $\objB$ on each dataset.
For all competitors, we report the results at the iteration when the cost function $\objB$ was minimized.

\subsection{Q1. Search Quality \& Speed} \label{sec:exp:quality}

In this subsection, we compare the considered algorithms in terms of speed and the quality of output near bi-cliques.

\smallsection{Evaluation metric}
For the quality, we use the total cost $\objB$ relative to the encoding cost of the input dynamic graph, i.e., 
\begin{equation}
	\text{Relative Cost}(\blockset, \GG) := \frac{\objB}{|\E|\cdot\LE} \label{eqn:cr},
\end{equation} 
where the denominator is the number of bits to individually encode every edge in the input graph $\GG$.
Since the denominator is a constant for a given $\GG$, Eq.~\eqref{eqn:cr} is proportional to the total cost $\objB$. Thus, as explained in Section~\ref{sec:problem:quality}, the lower Eq.~\eqref{eqn:cr} is, the more concise, precise, and exhaustive  detected near bi-cliques are.


\smallsection{Summary of comparison}
As seen in Figure~\ref{fig:quality}, \method gave near bi-cliques with the highest quality in all considered datasets. Specifically, the near bi-cliques detected by \method had up to $\mathbf{51.2\%}$ \textbf{better quality} in terms of the relative cost. 
Moreover, \method was up to $\mathbf{68.8\times}$ \textbf{faster} than \timecrunch, which detected near bi-cliques with the second highest quality in most cases.

\begin{figure}[t]
    \vspace{-3mm}
    \centering
	\includegraphics[width=0.9\linewidth]{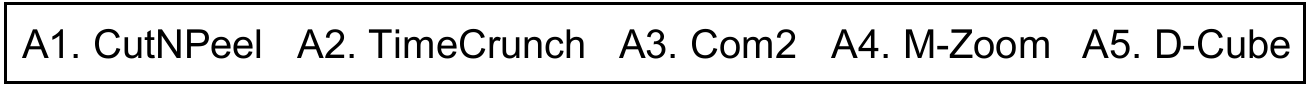} \\
	\vspace{-4mm}
	\subfloat[\small Enron]{\includegraphics[width=0.152\textwidth]{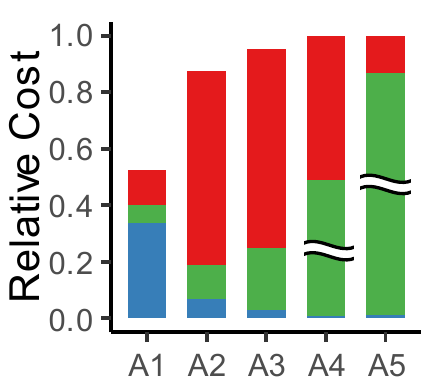}\label{subfig:stacked_enron}}
    \subfloat[\small DBLP]{\includegraphics[width=0.152\textwidth]{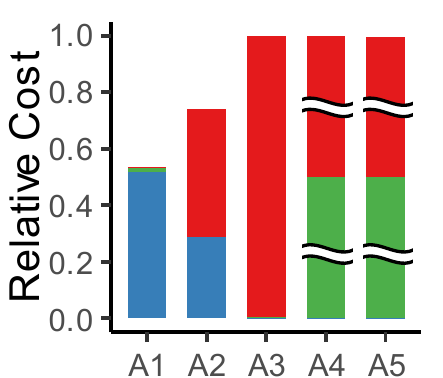}\label{subfig:stacked_DBLP}}
    \includegraphics[width=0.165\textwidth]{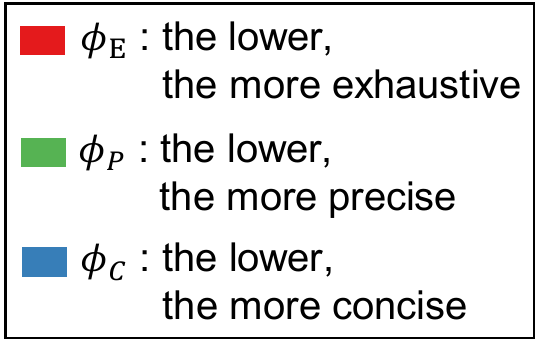} \\
    \vspace{-1mm}
    \caption{\label{fig:cost_portion} 
    \textbf{\method provides near bi-cliques with the best quality, and they are especially superior in exhaustiveness and preciseness.}
    See Appendix~B \cite{onlinesupplementary} for full results. }
\end{figure}

\smallsection{Detailed analysis}
In Figure~\ref{fig:cost_portion}, we divide the relative cost in Eq.~\eqref{eqn:cr} into three portions related to preciseness (i.e., $\objF_{P}(\correctionM)$ in Eq.~\eqref{eq:preciseness}), exhaustiveness (i.e., $\objF_{E}(\correctionP)$ in Eq.~\eqref{eq:exhaustiveness}), and conciseness (i.e., $\objF_{C}(\blockset)$ in Eq.~\eqref{eq:conciseness}), respectively.
The total cost was lowest in \method, and the near bi-cliques detected by it  were especially superior in exhaustiveness and preciseness.
While $\objF_{C}(\blockset)$ in \method was relatively high with many near bi-cliques, 
individual near bi-cliques detected by \method were precise with few missing edges, as shown in Figure~\ref{fig:patterns}, where we compare the size and preciseness of the near bi-cliques detected by the considered algorithms.
Noticeably, the near bi-cliques detected by \method are precise with extremely small ratios of missing edges, and their count is tractable and especially much smaller than the number of edges in the input graph.
\mzoom and \dcube tended to detect fewer larger subgraphs with higher ratios of missing edges than the other algorithms.

\subsection{Q2. Scalability} \label{sec:exp:scalability}
In this subsection, we test the scalability of \method. To this end, we created Erd\H{o}s-R\'enyi random graphs with various sizes and measured the running time of \method on them. The number of objects of each type was $0.1\%$ of the number of edges, and the largest graph had $2^{27}=134,217,728$ edges. 
As seen in Figure~\ref{fig:highlights}b in Section~\ref{sec:intro}, the running time of \method scaled near linearly with the number of edges and objects in the input dynamic graph. 

\subsection{Q3. Application: Compression} \label{sec:exp:compression}

In this subsection, we show a successful application of \method to lossless graph compression. We consider \comtwo and \timecrunch as competitors, which are, to the best of our knowledge, the only lossless compression methods for the entire history (rather than the current snapshot \cite{ko2020incremental}) of a dynamic graph.

The relative cost in Eq.~\eqref{eqn:cr} is the ratio of the number of bits for encoding the input graph losslessly using detected near bi-cliques and the number of bits for encoding each edge separately. Thus, it can naturally be interpreted as the compression rate.
Alternatively, the input graph can be encoded using near bi-cliques as suggested in \cite{araujo2014com2}, and the compression rate can be computed accordingly.
Thus, we measured it and Eq.~\eqref{eqn:cr} for all considered methods.
Additionally, for \timecrunch, we measured the compression rate in \cite{shah2015timecrunch}, which is based on an encoding method not applicable to the others.

As seen in Table~\ref{Tab:compress},
\method achieved the best compression in all considered datasets. Specifically, \method achieved up to $64.9\%$ better compression rates than the second best method.

\smallsection{Extra experiments} In the supplementary document \cite{onlinesupplementary}, we review extra experiments for Q4-Q6:
\begin{itemize}[leftmargin=9pt]
    \item{\textbf{Q4. Ablation Study:}} How much do adaptive thresholds  and partitioning contribute to the performance of \method?
    \item {\textbf{Q5. Parameter Analysis:}} How do the decrement rate $\decrement$ and the iteration number $\iter$ affect the performance of \method?
    \item{\textbf{Q6. Application 2 - Pattern Discovery:}} What interesting patterns does $\method$ detect in real-world data?
\end{itemize}

\section{Related Work}
\label{sec:related}

In this section, we review studies of finding dense (near) (bi-)cliques in static and dynamic graphs. 

\subsection{Finding (near) (bi-)cliques in static graphs}

Below, we focus on finding dense subgraphs in static graphs. 


\smallsection{Enumerating exact (bi-)cliques}
Finding (bi-)cliques (i.e., complete subgraphs), especially, enumerating all maximal (bi-)cliques, which are not a subset of any other clique, has been extensively studied \cite{makino2004new,eppstein2010listing,tsukiyama1977new,chiba1985arboricity,alexe2004consensus,liu2006efficient,kloster2019mining,dias2005generating,quine1955way}.
The problem of enumerating all maximal (bi-)cliques was shown to be NP-hard \cite{moon1965cliques,eppstein1994arboricity}. 

\smallsection{Enumerating near (bi-)cliques}
Considerable  attention has been paid to near (bi-)cliques with constraints on minimum degree \cite{batagelj2003m}, edge density \cite{abello2002massive,uno2010efficient}, and diameter \cite{balasundaram2011clique,moradi2018finding}.
Notably, $k$-cores \cite{batagelj2003m} for all $k$ can be found in linear time \cite{batagelj2003m}, while enumerating (maximal) quasi-cliques \cite{uno2010efficient}, (maximal) k-clubs \cite{moradi2018finding}, or (maximal) k-plexes \cite{mcclosky2012combinatorial,zhou2020enumerating} is NP-hard.
Similarly, near bi-cliques include a subgraph where (a) every node in one node set is adjacent to at least  $(1-\epsilon)$ of the nodes in the other node set \cite{mishra2004new}, (b) every node in both node sets is adjacent to at least $(1-\epsilon)$ of the nodes in the counterpart node set \cite{bu2003topological}, 
and (c) every node in both node sets is not adjacent to at most $\epsilon$ nodes in the counterpart node set \cite{sim2009mining}.
Regardless of the definitions, enumerating (maximal) near bi-cliques is NP-hard since their number can be exponential in the number of nodes.

\smallsection{Finding a partial set of near (bi-)cliques}
There also have been extensive studies on finding one or a predefined number of near (bi-)cliques.
A representative example is to find a subgraph that maximizes average degree \cite{goldberg1984finding,khuller2009finding,charikar2000greedy} or the ratio between the number of contained (bi-)cliques and the number of nodes \cite{tsourakakis2015k, mitzenmacher2015scalable}.
\vog \cite{koutra2014vog} searches for a partial set of near (bi-)cliques and chain subgraphs by which the input graph is summarized.


\subsection{Finding near bi-cliques in dynamic graphs}
Below, we focus on finding dense subgraphs and their occurrences over time in dynamic graphs.

\smallsection{Finding a predefined number of near bi-cliques}
Shin et al.~\cite{shin2018fast,shin2021detecting} and Jiang et al.~\cite{jiang2015general} studied the problem of finding a given number of dense subtensors~\cite{shin2018fast, shin2021detecting, jiang2015general} in a tensor (i.e., a multidimensional array).
For example, starting from the entire tensor, \mzoom~\cite{shin2018fast} and \dcube~\cite{shin2021detecting} greedily remove entries so that one of the proposed density measures is minimized. 
Since a dynamic graph is naturally expressed as a 3-way tensor, 
the above algorithms can be used for identifying dense subgraphs in a dynamic graph.

\smallsection{Summarizing dynamic graphs using near bi-cliques}
\comtwo \cite{araujo2014com2} and \timecrunch~\cite{shah2015timecrunch} aim to concisely describe the input dynamic graph using dense subgraphs, including near bi-cliques.
\comtwo uses rank-1 CP deomposition to obtain the score (i.e., corresponding value of the factor matrices) of each object and 
sorts the objects of each type based on the score.
Objects are added one by one to a near bi-clique greedily until the description length does not decrease, and the above step is repeated for detecting multiple near bi-cliques. 
\timecrunch 
aims to identify (near) (bi-)cliques and chain subgraphs.
To this end, \timecrunch decomposes the input graph at each timestamp into such subgraphs using \slashburn~\cite{lim2014slashburn}, and ``stitch'' some of the found ones over timestamps based on the description length.
Among these candidates, some are selected using multiple heuristics.



\begin{table}[t]
     \vspace{-3mm}
    \centering
    \caption{\label{Tab:compress} \textbf{
    \method consistently achieved the best lossless compression.} o.o.t.: out of time ($> 6$ hours). See Appendix~C \cite{onlinesupplementary} for full results with standard deviations.} 
    \scalebox{0.79}{
        \begin{tabular}{c|c|c|c|c|c|c|c}
            \toprule
            \multirow{2}{*}{\textbf{Metric}*} & \multirow{2}{*}{\textbf{Method}} & \multicolumn{6}{|c}{\textbf{Compression Rates in \%} (the lower, the better)} \\
            \cmidrule(lr){3-8}
            & & Enron & Darpa & DDoS & DBLP & Yelp & Weeplaces\\
            \midrule
            \multirow{3}{*}{Eq.~\eqref{eqn:cr}} & \method & \underline{52.5} & \underline{32.3} & \underline{13.1} & \underline{53.7} & \textbf{77.1} 
            & \underline{55.9}\\
            & \timecrunch & 87.7 & 37.2 & 31.5 & 74.2 & 100.6 & o.o.t. \\
            & \comtwo & 95.5 & 100 & o.o.t. & 99.9 & o.o.t. & o.o.t.\\
            \midrule
            \cite{shah2015timecrunch}** 
            & \timecrunch & 86.3 & 36.7 & 16.3 & 79.7 & 100.0 & o.o.t. \\
            \midrule
            \multirow{3}{*}{\cite{araujo2014com2}} & \method & \textbf{48.3} & \textbf{27.1} & \textbf{8.0} & \textbf{52.4} & \underline{78.0} & \textbf{52.4}\\
            & \timecrunch & 79.3 & 34.6 & 22.8 & 63.9 & 100.2 & o.o.t. \\
            & \comtwo & 58.3 & 202.8 & o.o.t. & 57.5 & o.o.t. & o.o.t.\\
            \bottomrule
            \multicolumn{8}{l}{* Different metrics are based on different encoding methods.}\\
            \multicolumn{8}{l}{** The encoding method used in \cite{shah2015timecrunch} is not  applicable to \method and \comtwo.}\\
        \end{tabular}
    }
\end{table}

\section{Conclusion}
\label{sec:conclusion}

In this work, we consider the problem of finding a concise, precise, and exhaustive set of near bi-cliques in a dynamic graph.
We formulate the problem as an optimization problem whose objective combines the three aspects (i.e., conciseness, preciseness, and exhaustiveness) in a systematic way based on the MDL principle.
Our algorithmic contribution is to design \method for the problem.
Compared to a widely-used top-down greedy search, \method reduces the search space and at the same time improves search accuracy, through a novel adaptive re-partitioning scheme.
We summarize the strengths of \method as follows:
\begin{itemize}[leftmargin=9pt]
    \item \textbf{High Quality}: Compared to its best competitors, \method finds near bi-cliques of up to $\mathit{51.2\%}$ \textit{better quality} (Figures~\ref{fig:quality}-\ref{fig:patterns}).
    \item \textbf{Speed}: \method is up to $\mathit{68.8\times}$ \textit{faster} than the competitor with the second best quality (Figure~\ref{fig:quality}).
    \item \textbf{Scalability}: \method scales to graphs with up to $\mathit{134}$ \textit{million edges}, near-linearly with the size of the input graph (Figure~\ref{fig:highlights}b).
    \item \textbf{Applicability}: Using \method, we achieve up to $\mathit{64.9\%}$ \textit{better compression} than the best competitor (Table~\ref{Tab:compress}), and we spot interesting patterns (e.g., relevant conferences in Figure~\ref{fig:highlights}c).
\end{itemize}

{\small \smallsection{Acknowledgements} This work was supported by Samsung Electronics Co., Ltd., National Research Foundation of Korea (NRF) grant funded by the Korea government (MSIT) (No. NRF-2020R1C1C1008296),
and Institute of Information \& Communications Technology Planning \& Evaluation (IITP) grant funded by the Korea government (MSIT) (No. 2019-0-00075, Artificial Intelligence Graduate School Program (KAIST)).}




\vfill\eject
\bibliographystyle{ACM-Reference-Format}
\balance
\bibliography{reference}



\end{document}